\documentclass[showpacs,twocolumn,aps,prl]{revtex4}
\usepackage{amssymb}

\usepackage{graphicx}
\usepackage{dcolumn}
\usepackage{bm}

\def\be{\begin{equation}}
\def\ee{\end{equation}}
\def\bea{\begin{eqnarray}}
\def\eea{\end{eqnarray}}
\def\ba{\begin{array}}
\def\ea{\end{array}}

\def\0{$\Gamma_0$}

\begin{document}

\title{Negative refraction and plano-concave lens focusing in one-dimensional photonic crystals}

\author{P. Vodo}
\author{W. T. Lu}
\email{w.lu@neu.edu}
\author{Y. Huang} 
\author{S. Sridhar}
\email{s.sridhar@neu.edu}

\affiliation{Department of Physics and Electronic Materials Research
Institute, Northeastern University, Boston, Massachusetts 02115}

\date{\today }

\begin{abstract}
Negative refraction is demonstrated in one-dimensional (1D) dielectric photonic crystals (PCs)
at microwave frequencies. Focusing by plano-concave lens made of 1D PC due to negative refraction 
is also demonstrated. The frequency-dependent negative refractive indices,
calculated from the experimental data matches very well with those determined 
from band structure calculations. 
The easy fabrication of one-dimensional photonic crystals may open the door for new applications.
\end{abstract}

%\pacs{03.65.Sq}
\maketitle

Negative refraction \cite{Veselago} and related phenomena such as 
flat lens imaging \cite{Pendry,Luo02,Parimi03} and plano-concave 
lens focusing \cite{Vodo05}
have attracted a lot of attention in physics and engineering.
Negative refraction allows subwavelength imaging \cite{Luo03} and 
focusing of far field radiation by concave 
rather than convex surfaces with the advantage of reduced 
aberration \cite{Schurig,Chen06} for the same radius of curvature. 
Negative refraction has been realized in two- and three-dimensional structures in 
metamaterials \cite{Shelby,Parazzoli} and photonic crystals \cite{Parimi04,LuZ}.
But negative refraction has not yet been demonstrated in 
one-dimensional (1D) photonic crystals (PCs).

In this letter we present a study of left-handed electromagnetism
in 1D PCs at microwave frequencies. Negative refraction is achieved 
in the second band of the 1D PCs. 
Focusing of plane wave radiation 
by plano-concave lenses made of 1D PC is also demonstrated. 
The inverse experiment, in which the lens 
produces plane waves from a point source placed at the focal length, 
at the same frequency of operation, is confirmed as well.
The frequency-dependent negative refractive indices, calculated from 
the experimental data matches very well with those determined from 
band structure calculations. 
The easy fabrication of 1D PCs may open the door for new applications.

The experiments were carried out in a parallel-plate waveguide of 
height $h=1.25$ cm and size $3\times6$ ft$^2$.
For frequency below 12 GHz, the excitation in these quasi-2D 
system is the transverse magnetic (TM) modes with the electric field 
in the vertical direction.
The electric field of the microwaves is scanned using a monopole antenna 
attached to a X-Y robot in the frequency window 3-11.5 GHz. 
An HP-8510C network analyzer 
is used for measuring the transmission characteristics. 
A schematic diagram of the experimental setup is shown in Fig. 1.

Alumina bars with permittivity $\epsilon=8.9$ were laid out to 
form prisms of right angle triangles. 
The bars have a height $h=1.25$ cm and width $d=0.5$ cm. 
The refraction experiments were performed on two PC prisms.  
The first prism (PC1) has lattice constant $a=1$ cm and the incident angle 45$^\circ$ while the 
second one (PC2) has $a=0.8$ cm and the angle 51$^\circ$.
All bars have a perpendicular cut instead of a slanted one,
as numerical simulations \cite{Lu06} show that a perpendicular cut 
 reduces the modulations of the outgoing waves, 
partly due to the absence of sharp corners.
A plane wave incident normally to one surface is 
refracted by the hypotenuse of the 1D PC prism as shown in Fig. 1, 
defined as the surface of refraction. 

\begin{figure}
%\centerbmp{8.5cm}{5.5cm}{NR_1D_fig1.jpg}
\center{\includegraphics [angle=0,height=5.5cm]{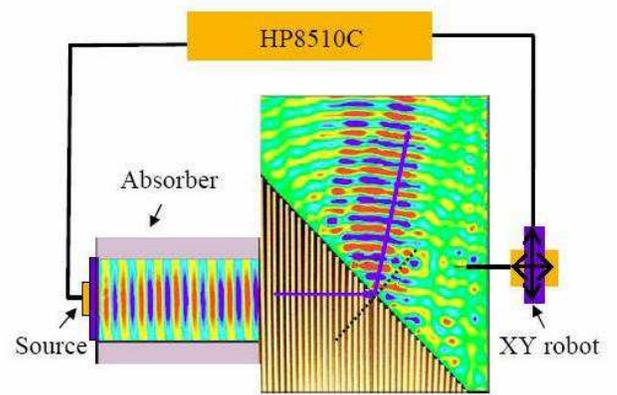}}
\caption{Schematic diagram of the microwave experimental setup
and negative refraction of 
plane waves by the 1D PC1 prism. Angle of incidence is 
45$^\circ$ and angle of refraction is –36.9$^\circ$ resulting 
in negative refraction with refractive index 
$n_{p}=-0.85$ at 10.55 GHz. 
The real part of $S_{21}$ scale: on the left side varies from -0.015  
to 0.015, on the right side from -0.006  to 0.006.}
\label{fig1}
\end{figure}

The 1D PC is a model that is exactly soluble \cite{Joannopoulos}. 
For the TM modes, it is just the Kronig-Penny model \cite{Kronig} with
an energy dependent potential. 
The band structure for the TM modes for the filling factor $d/a=0.5$ 
is shown in Fig. 2.
A band gap is located between 5.55-8.9 GHz. 
The second pass band is between 8.9-12.7 GHz and 
has negative refractive index. 
The refraction of a microwave beam by the PC1 prism at 10.55 GHz is shown in Fig. 1.
By fitting the outgoing beam with a plane wave, the wave front 
is determined and an effective index $n_p=-0.85$ is obtained
using Snell's law.

Although positive refraction is predicted for both PC prisms
in the first band, the incident angle for each PC prism exceeds 
the corresponding Brewster angle, 
resulting in total internal refraction. 
The average power of the scanned points is 
plotted as a function of frequency in Fig. 2. 
While the outgoing signal is very weak for the total internal
reflection and band gap frequency regions,
an interesting observation is that there 
is a strong leaking from the prism near the edges of the band gap.

\begin{figure}
\center{\includegraphics [angle=0,height=5.5cm]{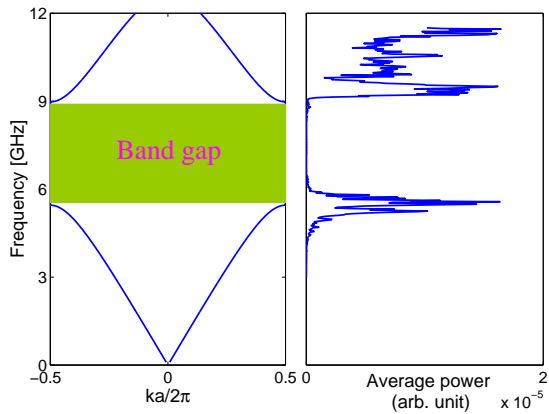}}
\caption{(a) Band structure of the TM modes of the 1D PC with $a=1$ cm,
$d=0.5$ cm, and $\epsilon=8.9$.  
(b) Frequency versus average outgoing power, calculated as the average 
of $|S_{21}|^2$ in the scanned area.
From 3-5 GHz all energy in internally reflected. 
Note the leaking of energy from around the edges of the band gap.
Measurement was performed between 3-11.5 GHz.}
\label{fig2}
\end{figure}

The same alumina bars were used to form a plano-concave lens of 1D PC.
The concave radius of the lens is $R=18$ cm.
A sharp focal point is located at 6.15 cm away from the curved surface
when a microwave beam incidents at frequency 9.5 GHz. 
From left to right in Fig. 3, the incoming plane wave, a real picture 
of the PhC lens and the emerging mapped field are shown. 
Clear focusing is observed in the frequency range 9.2-11.5 GHz. 

\begin{figure}
%\centerbmp{7cm}{6.03cm}{NR_1D_fig3.jpg}
\center{\includegraphics [angle=0,height=5.5cm]{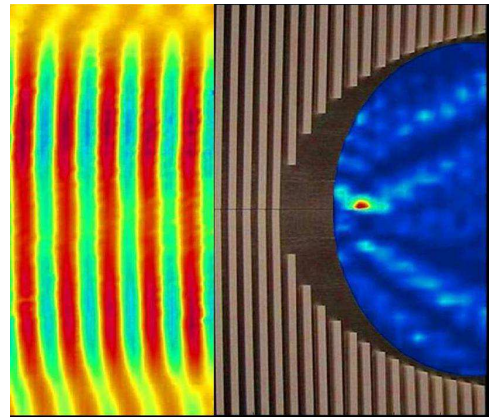}}
\caption{Focusing by a 1D plano-concave PC lens having radius 
of curvature $R=18$ cm. The focus point observed at 9.5 GHz is  
6.15 cm from the concave lens surface. A photograph of the 
PC is superimposed to obtain the final figure. 
On the left side, field map of the incoming plane wave is 
shown (real part of transmission coefficient) and on the 
right side, intensity  of the focus point. Scale: 
on the left, from -0.03 to  0.03, on the right side from 0  to 
$4\times10^{-3}$.  Dimensions of the lens are  $19\times 38$ cm$^2$.}
\label{fig3}
\end{figure}

An inverse experiment in which a point source is 
placed at the observed focal point of the lens at a single frequency
is also carried out. 
As shown in Fig. 4, a circular wave front from the point 
source after passing through the lens emerges as a plane wave. 
These two remarkable results validate the behavior of a left-handed 
plano-concave lens.

\begin{figure}
%\centerbmp{7cm}{5.83cm}{NR_1D_fig4.jpg}
\center{\includegraphics [angle=0,height=5.5cm]{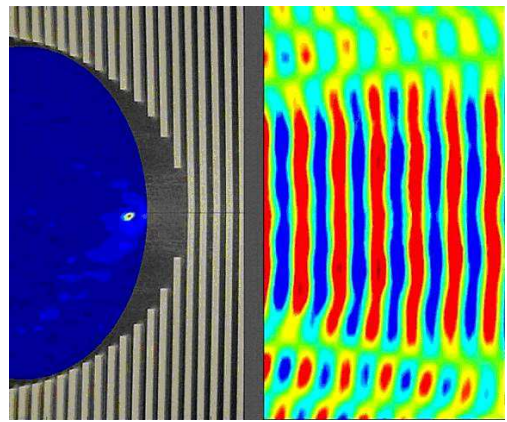}}
\caption{Field maps of the incident source and the emerging plane wave. 
Scale: on the left side intensity varies from  0 to 0.14, 
on the right side the real part of $S_{21}$ from  -0.03 to 0.02. 
The Source is placed at the focal length 6.15 cm and plane wave is observed at 9.55 GHz.}
\label{fig4}
\end{figure}

The refractive index of the lens can be estimated using the 
lens equation $n=1-R/f$ which is valid for thin plano-concave and 
plano-convex lenses in the geometric optics limit.
Here $R$ is the radius of curvature and $f$ is the focal length. 
Using this equation we get $n=-0.85$ at 9.5 GHz for the lens shown in Fig. 3. 
A real focus by a plano-convex lens is achieved 
with $n>1$ and $R<0$ while for the plano-concave lens 
with $n<1$ and $R>0$. 

The refractive indices $n_p$ determined from the prism refraction experiments
(PC1 and PC2) and the plano-concave lens one are shown in Fig. 5. 
Very good agreement with those calculated from the band structure is observed.
The index $n_p$ determined from the focusing experiment 
fits better with theoretical results as the frequency is increased. 
This may be due to the reduced finite-size effect and aberration at higher frequency. 

\begin{figure}
\center{\includegraphics [angle=0,height=6.5cm]{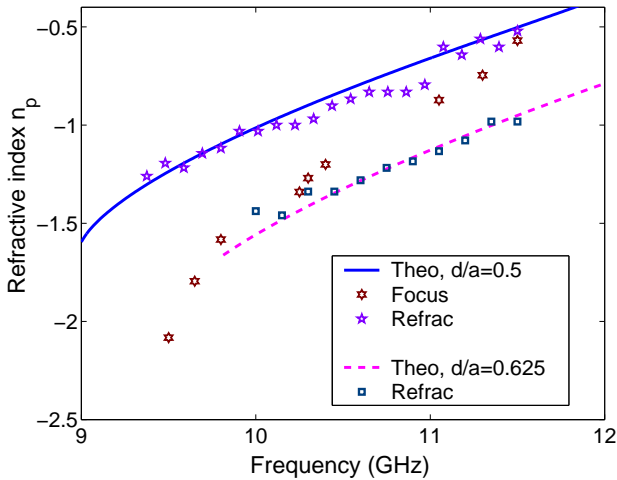}}
\caption{Refractive indices $n_p$ determined from the focusing and refraction 
experiments for the TM modes of the 1D dielectric PCs with $\epsilon=8.9$. 
The solid line is for PC1 ($d/a=0.5$) and the dashed line for PC2 ($d/a=0.625$).} 
\label{fig5}
\end{figure}

The nature of the left-handed electromagnetism and focusing can 
be understood from the dispersion characteristics of the 1D PC. 
From the band structure shown in Fig. 2, it 
can be deduced that in the second band the wave vector  is 
in opposite direction to group velocity,
${\bf v}_g\cdot {\bf k}<0$, resulting in negative refraction 
in the second band and correspondingly negative refractive indices \cite{Parimi04}. 

Bandwidth for obtaining a sharp focus point is a crucial parameter 
for applications of the left-handed lenses. 
Due to the resonant nature of the metamaterial the bandwidth is 
usually restricted to a narrow region and the dispersion 
is strong \cite{Parazzoli04}.  
The PC1 reveals a wide bandwidth of 3.8 GHz, which is 35\% 
at the current operating frequencies. The weaker dispersion in the PC 
makes it a better candidate for 
focusing a pulse or broadband radiation.  

The present PC lens with negative refraction has several 
advantages when compared to the one with positive refraction. 
Lenses with reduced geometric aberrations produce sharper image 
with enhanced resolution and find numerous applications. 
Larger radius of curvature gives the advantage of reduced aberration 
in the image formed. A PC lens having the same focal length 
as that of a conventional lens weighs far less, and is attractive to 
space applications. The tailor made refractive index achievable in 
PC materials \cite{Vodo04} allows further control on the focal length and thereby 
helps to reduce the length of the optical systems. 

In conclusion the feasibility of designing a 1D broadband left-handed 
PC lens is experimentally demonstrated. Negative refraction of plane waves 
and plano-concave lens focusing is achieved by 1D PCs. 
The focal length follows the standard laws of geometrical optics 
combined with negative refraction. The measured values of 
refractive indices of the lens are in excellent agreement with 
those determined from band structure calculations from both 
refraction and focusing experiments. 
Earlier work has shown that 1D PCs can be used as omnidirectional
reflectors \cite{Fink,Ibanescu}. 
The observed negative refraction in 1D PC reported here, adds 
a new feature to these simple systems.

We thank E. Brownell for assistance with the experiments.
This work was supported by the Air Force Research Laboratories, 
Hanscom AFB and NSF grant No.
PHY-0457002.

\end{document}